# Rheological properties of brushes on cancerous epithelial cells under the influence of an external oscillatory force

J. D. Hernández Velázquez[1], S. Mejía – Rosales[1], A. Gama Goicochea[2*]

[1]Centro de Investigación en Ciencias Físico – Matemáticas, Facultad de Ciencias Físico - Matemáticas, Universidad Autónoma de Nuevo León, San Nicolás de los Garza, Nuevo León 66450, Mexico

[2]Tecnológico de Estudios Superiores de Ecatepec, Av. Tecnológico s/n, Ecatepec, Estado de México 55210, Mexico

## ABSTRACT

The rheological properties of brushes of different length on the surface of human epithelial cancerous cells are studied here by means of coarse – grained numerical simulations, where the surface of the cell is subjected to an external oscillatory force acting on the plane of the cell's surface. We model explicitly the tip of an atomic force microscope and the cancerous cell as a surface covered by brushes of different length, and take into account the interactions of the brush chains with the tip and with each other, leading to complex rheological behavior as displayed by the profiles of viscosity and the friction coefficient of this complex system. We comment briefly on how these findings can help in the experimental effort to understand the nature of the cancer growth in human epithelial cells.

[*] Corresponding author. Electronic mail: agama@alumi.stanford.edu

# I Introduction

It is known that cervical cancer is one of the leading types of cancer in women over 35 years of age [1]. In an effort to understand the characteristics of this and other types of cancer, several physical approaches have been applied recently to gain new insights about how metastasis occurs, development of non – invasive detection methods, and perhaps even help in the design of new treatment [2-5]. One of those approaches consists in the characterization of cancerous cells using atomic force microscopy (AFM) [6]. Using AFM, several groups have determined that cancerous cells from different types of tissue are softer than their normal counterparts [2-5, 7, 8], while others claim that the mechanical response of malignant cells depends on the type of cancer [9, 10]. Most of the research on this topic has focused on the response of the cells' surface as a whole. One notable exception is the work of Iyer and collaborators [11], who measured the force exerted by the tip of an AFM on the *brushes* that covered cancerous cervical epithelial cells, finding a force profile qualitatively different from the one measured on healthy cells. The difference was attributed to the inhomogeneous composition of the brushes that cover the cancer cells, while normal cells are covered by brushes of approximately the same length. These experiments are important, not only because they help establish the role played by the brushes that typically cover most types of epithelial cells, which may be different from the role played by the surface of the cell, but also because they can be used to design improved detection methods.

On the modeling side, it has been shown [12] that the softness or stiffness of the individual "structures" that make up the brushes is responsible of the apparent softness or stiffness of the brush as a whole. These "structures" or "molecules" can be of different nature; they can be microridges, microtubules or microvilli [13], whose purpose is to provide motility to the cell and increase nutrient absorption. They can move on the surface of the cells, although it has been shown [12] that the mechanical response of epithelial cells to the force exerted by an AFM is considerably more dependent on the softness/stiffness of the brushes than on their ability to move or not on the surface of the cells. Since it can be envisaged that a correlation exists between the stiffness of brushes on cancerous cells, and cancer stadia, it is befitting to study this issue further.

In this contribution we report results of non – equilibrium numerical simulations at the mesoscopic level of brushes on a surface under oscillatory flow on the plane of the surface. The motivation for performing simulations of this type stems from recent experiments with AFM on cancer cells under non – equilibrium conditions [11]. Two key aspects are novel in this work: firstly, the curvature of the tip of the AFM is incorporated explicitly, which makes our predictions useful for those workers who use nanometer – size tips in their AFM. Secondly, we have included a three – body interaction between neighboring bonds along the chains that make up the "molecules" of the brushes, so that their softness or stiffness can be controlled directly. By subjecting this system to a sinusoidal external force along the plane of the cell's surface one can obtain rheological properties such as the viscosity and friction between the brush and the AFM, which can be useful for an improved characterization of this illness. The rest of this chapter is organized as follows: in section II we presented the model, methods and systems studied; the results and their discussion are reported in section III. Finally, our conclusions are laid out in section IV.

## II Models and Methods

The simulations whose results are the purpose of this contribution use the interaction model known as dissipative particle dynamics (DPD) [14], which consists of the simultaneous integration of the equation of motion of all particles that make up the system, to obtain their positions and momenta. In this regard, DPD is identical to standard simulations of molecular dynamics [15]. The difference stems from the introduction of a natural thermostat into the algorithm, which arises from the balancing of the dissipative and random forces [16]; this is the major advantage of DPD over other simulation algorithms. The conservative force that acts between particles is simple (repulsive, linearly decaying) and, like the random and dissipative forces, of short range, which is the reason why DPD is useful to model systems at the mesoscopic level. Figure 1 illustrates the coarse – graining degree and the nature of the conservative force in the DPD model. However, it must be kept in mind that one is free to choose other types of forces, such as the Lennard – Jones model, with the DPD algorithm.

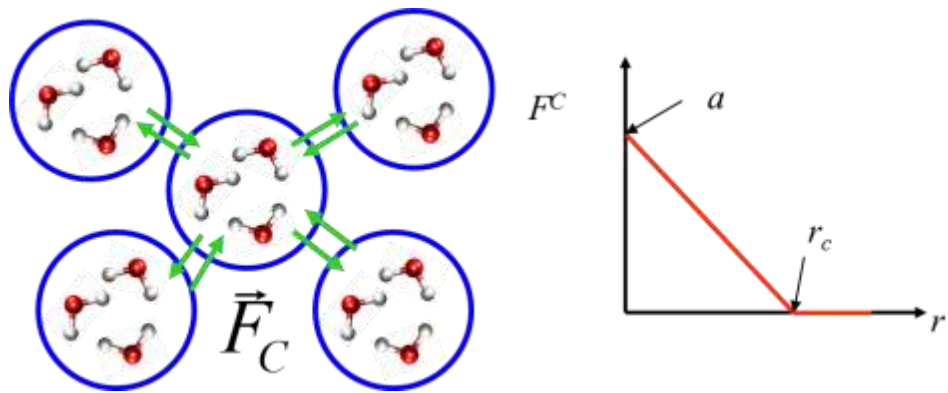

**Fig. 1** A schematic representation of the coarse – graining involved in DPD, where in this particular example each DPD particle (blue circles) groups three water molecules. The conservative force ($F_C$) acts as a local pressure between pairs of particles and it has a simple, linearly decaying and repulsive character –see the red line in the right panel of the figure. The conservative force becomes zero at relative distances larger than a cutoff radius, $r_c$, and it is equal to an interaction constant $a$ when $r=0$, which is determined by the chemical nature of the pair of DPD interacting particles. Adapted from [17].

As stated above, one of the major advantages of the DPD model is its naturally emerging thermostat, which has been shown to compete very favorably with other thermostats, especially under non – equilibrium circumstances [18], which are precisely the central focus of this work. Figure 2 is meant to show what the forces that make up the thermostat represent, namely, the dissipative force ($F_D$) and the random force ($F_R$), coupled through the fluctuation – dissipation theorem to lead to constant global temperature [16].

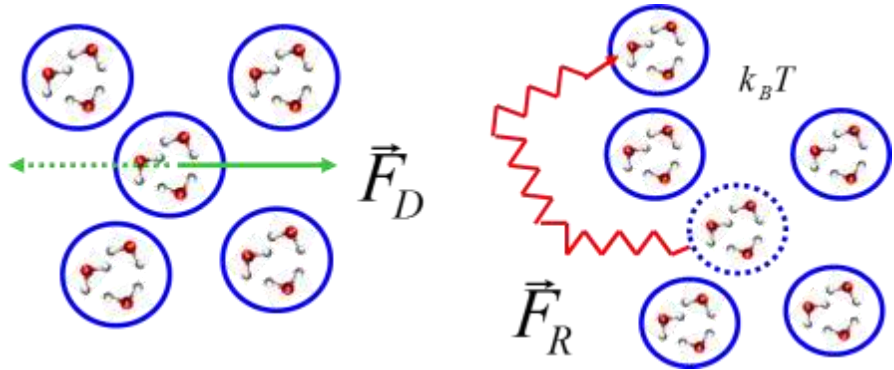

**Fig. 2** Illustrative representation of the forces that make up the thermostat in DPD; $F_D$ represents the dissipative force, which accounts for the local viscosity of the fluid, while $F_R$ is responsible for the local Brownian motion. The perfect coupling between these contributions keeps the global temperature fixed. Adapted from [19].

This model is now well known and there are recent reviews available [18, 20], therefore we shall skip details for the sake of brevity. We solve the forces for the system of particles using the algorithm of molecular dynamics, adapted to DPD [21].

Since our aim is to model experiments performed with AFM, we have incorporated explicitly its curved tip into the system, constructing it from individual DPD particles. Additionally, we model the brush – covered cancerous cell as a flat surface on top of which there are grafted chains of three different length, as suggested by recent experiments [11]. These linear chains are made up of DPD particles, joined by freely rotating harmonic springs, which interact with particles on the same chain and with those on other chains according to the DPD rules given by the forces illustrated in Figs. 1 and 2. In addition to those interactions we have introduced a three – body harmonic potential that acts between neighboring bonds, whose purpose is to introduce rigidity to the chains. There are also monomeric solvent particles that permeate the brush, which are meant to represent the buffer used in AFM experiments with cells *in vitro*. Lastly, the cell's surface on which the brushes are grafted is subjected to a sinusoidal force along the *x* – direction with maximum amplitude ($2/w$) and fixed frequency ($w=n\pi/100\Delta t$). The motivation for doing so arises from force measurements on oscillating polymer brushes using AFM [22, 23]. In Fig. 3 we present the system simulated in this work.

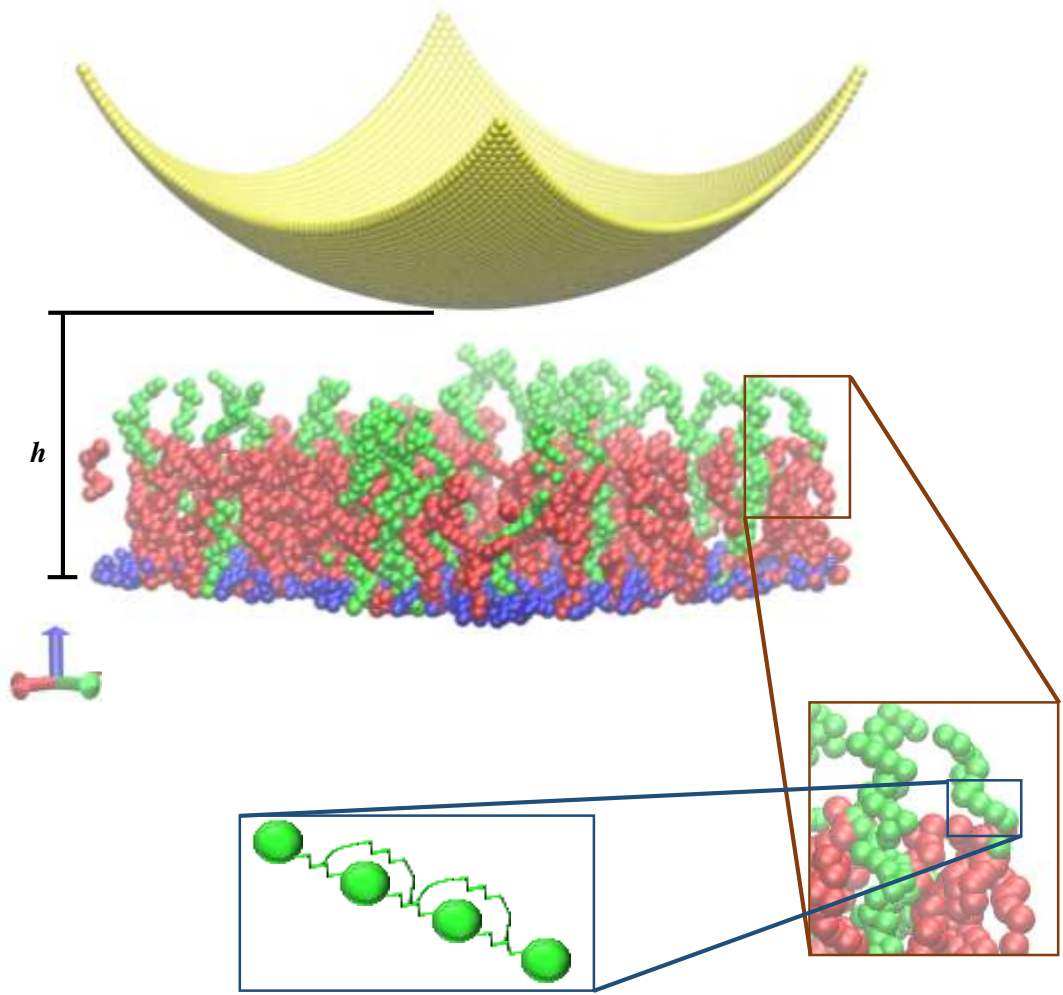

**Fig. 3** Model for brushes on cancerous cells probed by the tip of an AFM (in yellow) whose curvature radius is $R$. The brushes have three different lengths and are represented by chains of different colors, with the blue ones representing the smallest, followed by the red ones, and the Green chains are the largest. The beads that make up the chains are joined by harmonic springs and by angular harmonic springs (see insets). The number of beads that make up the chains ($N$), and the number of chains per unit area on the cell's surface ($\Gamma$) are $N_1 = 5, N_2 = 30, N_3 = 42$ and $\Gamma_1 = 1.76$ nm$^{-2}, \Gamma_2 = 0.49$ nm$^{-2}, \Gamma_3 = 0.20$ nm$^{-2}$. The subindexes 1, 2 and 3 refer to the blue, red and green chains, respectively. The cell lies on the $xy$ – plane and the brush is placed at a fixed distance, $h$.

As seen in Fig. 3, the cell's surface and the brushes grafted onto it lay on the $xy$ – plane, while the tip of the AFM exerts pressure along the $z$ – axis, with the tip of the AFM probe being placed at a fixed distance, $h$, along the $z$ – direction.

The interaction parameters $a_{ij}$ for the conservative force between the $i$-th and $j$-th beads depend on the coarse-grained degree, i.e., the number of molecules grouped into a DPD bead. For these simulations, we considered a coarse-grained degree equal to three, which leads to the parameters $a_{ij}$ shown in Table 1. We chose $\sigma = 3$ as the noise amplitude for the random force, and $\gamma = 4.5$ as the friction coefficient included in the dissipative force; the random and dissipative forces are coupled through the fluctuation-dissipation theorem, such that $\sigma^2/2\gamma = k_B T = 1$. The spring – like models for the distance- and angle-dependent interactions between neighboring beads (see the green beads in the inset of Fig. 3) are given by eqs. (1) and (2), respectively:

$$F_{spring} = k_s(r_{ij} - r_0)\hat{r}_{ij}, \qquad (1)$$

$$F_{angular} = k_a \sin(\theta_0 - \theta_{ijk}), \qquad (2)$$

where $k_s$ and $k_a$ are the constants of two-body and three-body spring forces, respectively, $r_0 = 0.7 r_c$ is the relaxation distance between two adjacent beads attached by a Hookean spring, $\theta_{ijk}$ is the angle between the bonds $\hat{r}_{ij}$ and $\hat{r}_{jk}$ formed by three adjacent beads (as if the bonds were attached by an angular spring) and $\theta_0 = 180°$ is the relaxation angle between these bonds. We chose the spring constants as $k_s = 100\ [k_BT/r_c^2]$ and $k_a = 100\ [k_BT/r_c]$, as those values have been successfully tested before [24].

The system conformed by the solvent and the brushes is confined by two walls (see Fig. 3); the bottom wall (located at $z = -l_z/2$, where $l_z$ is the length of the simulation box in the $z$ direction) is an implicit wall that represents the cell's surface, and its interaction with other beads is a linearly decaying short-range force, given by

$$\boldsymbol{F}_{wall}(z_i) = \begin{cases} a_{wi}(1 - z_i/z_c)\hat{\boldsymbol{z}} & z_i \leq z_c \\ 0 & z_i > z_c \end{cases}, \tag{3}$$

where $a_{wi}$ is the maximum interaction of the surface with the particle $i$ (see Table 1), $z_i$ is the distance of the particle $i$ to the surface, $z_c$ is the cutoff radius and $\hat{\boldsymbol{z}}$ is the unit vector in the z direction. The top wall (at $z = l_z/2$) is an explicit surface that represents an AFM tip, formed by a set of DPD beads arranged on a surface with a curvature radius $R = 0.8l_x$, where $l_x$ is the size of the box in the $x$ - direction (this surface is represented in Fig. 3 by yellow beads). The beads on this surface interact with the other particles on the same surface through their conservative DPD interaction, but their dissipative and random forces are zero, and those beads remain at rest. The conservative interaction between the AFM beads and those of the fluid (solvent and brush beads) is more repulsive than that between AFM beads, so that the tip remains impenetrable to the fluid. The full set of interaction parameters $a_{ij}$ is shown in Table 1.

The brushes attached to the cancerous cell are modeled by linear chains of beads of three different sizes: $N_1 = 5$, $N_2 = 30$ and $N_3 = 42$. We set the number of chains per unit area to $\Gamma_1 = 1.76 nm^{-2}$, $\Gamma_2 = 0.49 nm^{-2}$ and $\Gamma_3 = 0.20 nm^{-2}$ for the short, medium-sized, and large chains, respectively. The reason for choosing those values for the chains' length and grafting densities is because we want to model the relative differences in length and grafting density found in AFM experiments performed on cancerous human cervical cells [11]. In these simulations we introduced an oscillatory force along the $x$ – axis acting on the beads attached to the cell's surface (the *heads* of the brushes, see Fig. 4). This force is given by:

$$F_x(\Delta t) = A \cos(wn\Delta t)\hat{x},  \qquad (4)$$

where $w = \pi/100\Delta t$ is the oscillation frequency, $A = 2/w$ is the oscillation amplitude, $\Delta t$ is the time step, and *n* is the number of simulation steps.

| $a_{ij} \ [k_B T/r_c]$ | Solvent | Chain's head | Chain's tail | AFM beads | Cell's surface |
|---|---|---|---|---|---|
| **Solvent** | 78 | 79.3 | 79.3 | 140 | 100 |
| **Chain's head** | 79.3 | 78 | 78 | 140 | 60 |
| **Chain's tail** | 79.3 | 78 | 78 | 140 | 100 |
| **AFM beads** | 140 | 140 | 140 | 78 | 0* |
| **Cell's surface** | 100 | 60 | 100 | 0* | 0** |

**Table 1**. Table of all the interactions parameters $a_{ij}$ in the system (*since the distance between the AFM's tip and the cell's surface is larger than the cutoff radius. **Because it is an implicit surface).

The dimensions of simulation box are $l_x = l_y = 20r_c$ and $l_z = 26r_c$, the volume of the system is $V \approx 4933.77 r_c^3$ and the total density $\rho \approx 3$. The simulations proceed in two stages. First, we perform a thermalization process that consists of simulations of 5 blocks of $10^4$ time steps each, with a time step of $\Delta t = 0.001$. Once the system reaches thermal equilibrium, we carry out the production phase, with 10 blocks of $10^4$ time steps each; the time step selected for this phase is $\Delta t = 0.01$.

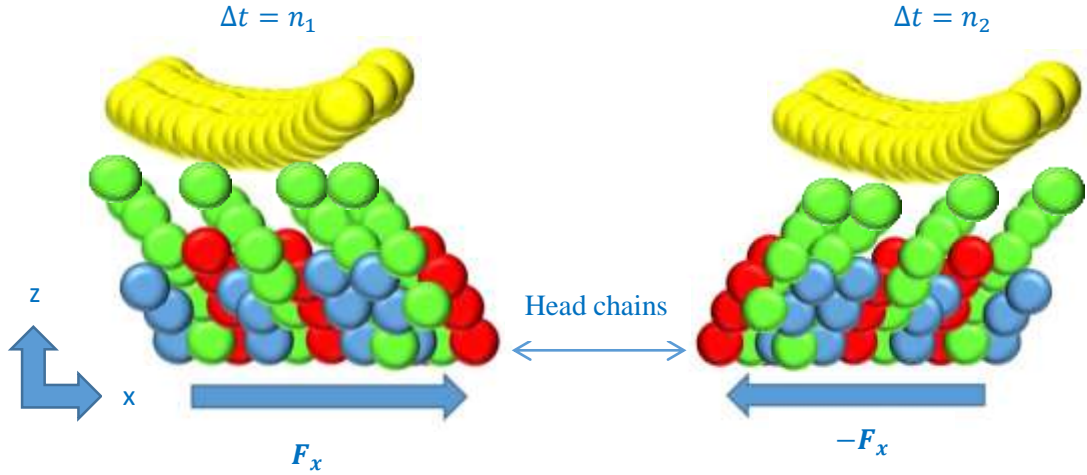

**Fig. 4** Schematic representation of the oscillatory force, eq. (4), acting on the chains' head beads "grafted" on the cell's surface. The oscillation period is $T = 200\Delta t$.

## III Results and Discussion

Let us start by considering the concentration profiles of the three types of brushes that make up the composite brush on the cancer cell's surface, when such brushes are moving under the influence of an external oscillatory force while being compressed by the tip of an AFM in the perpendicular direction. The results are shown in Fig. 5.

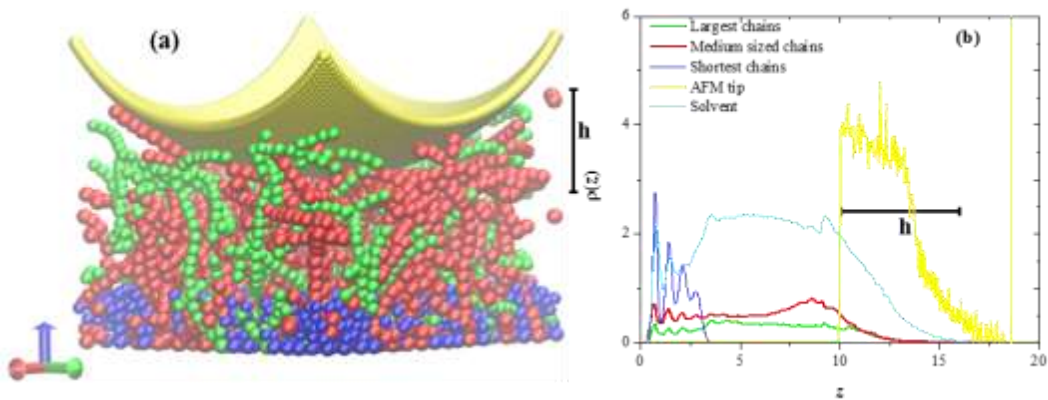

**Fig. 5** (a) Snapshot of the brushes on a cancer cell as they are being compressed by the tip of the AFM. Notice how some chains relieve compression by moving sideways. The solvent particles are omitted, for simplicity. (b) Concentration profiles of each type of brush, in reduced DPD units. The profile of the solvent (in cyan) and that of the tip of the AFM (in yellow) are shown also.

In the snapshot shown in Fig. 5(a), one recognizes that some of the largest chains, i.e. those represented in red and green, have moved away from the tip of the AFM to relieve some pressure. Figure 5(b) shows the concentration profiles of each type of brush; the blue curve corresponds to the shortest chains, the red and green curves are the profiles of the medium – sized and largest chains, respectively. The oscillations are related to the ordering of the beads that make up the chains close to the surface of the cell; notice also how the solvent particles penetrate the brushes all the way down to the surface of the cell. The period of the oscillations is roughly the size of the DPD particles, as expected [25]. There is a local maximum in the profile of the solvent right about where the profile of the smallest chains (blue line) goes to zero, which is due to the fact that the global density of the system is kept constant, therefore if there is a deficit in the brush profile it is compensated by the solvent's concentration. The same phenomenon occurs where the concentration of the medium – sized brushes decays, close to the tip of the AFM (about $z = 10$ in Fig. 5(b)). Lastly, it is quite remarkable how the profiles of the brushes follow the one of the AFM probe once the come into contact; see the profiles for $z > 10$ in Fig. 5(b). This aspect is the consequence of both the largest brushes and the solvent particles interacting with the curved side of the AFM probe and therefore that interaction is expected to be reflected in the rheological properties we shall calculate.

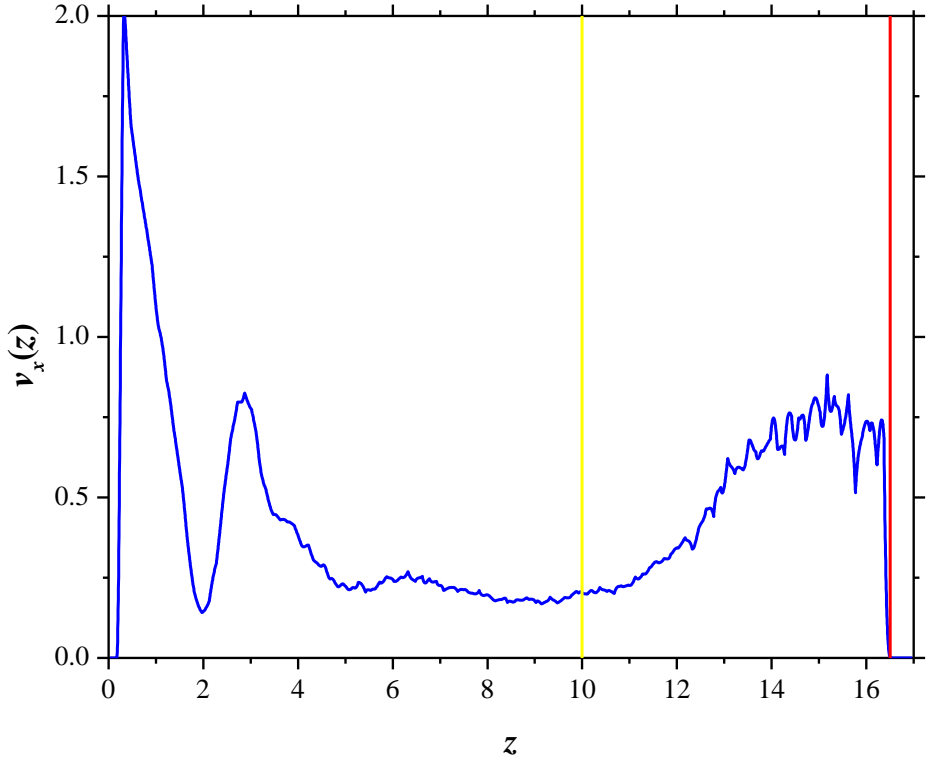

**Fig. 6** Profile of the $x$ – component of the velocity of the beads that make up the chains on the surface of the cancerous cell. Since the brushes are under the influence of an oscillatory external force, we used the maximum values of such component to obtain the averages in every slice used to make this profile. The vertical yellow and red lines indicate the position where the tip of the AFM probe is placed and where it ends, respectively. The scales on the axes are expressed in reduced DPD units.

In Fig. 6 we show the profile along the $z$ – direction of the $x$ – component of the velocity of the beads that make up the brushes shown in the previous figures. It must be recalled that there is an external oscillatory force applied to the surface of the cell along the $x$ – axis, which means that, to obtain the profile in Fig. 6 we chose the maximum values of $v_x(z)$ in every slice along the $z$ – direction. The oscillations close to the surface of the cell are the result of the collective motion of the brushes, which are relatively close to the surface to respond to the oscillations imposed by the external force at $z = 0$. Since those correlations tend to disappear as one moves away from the surface of the cell, the second maximum seen in Fig. 6 is of smaller amplitude than the first. Close to the tip of the AFM the profile is approximately constant on average, because the fluid is far enough from the oscillating surface, but once it comes into contact with the tip of the AFM (yellow vertical line in Fig. 6), the particles have more freedom to move and the velocity profile grows.

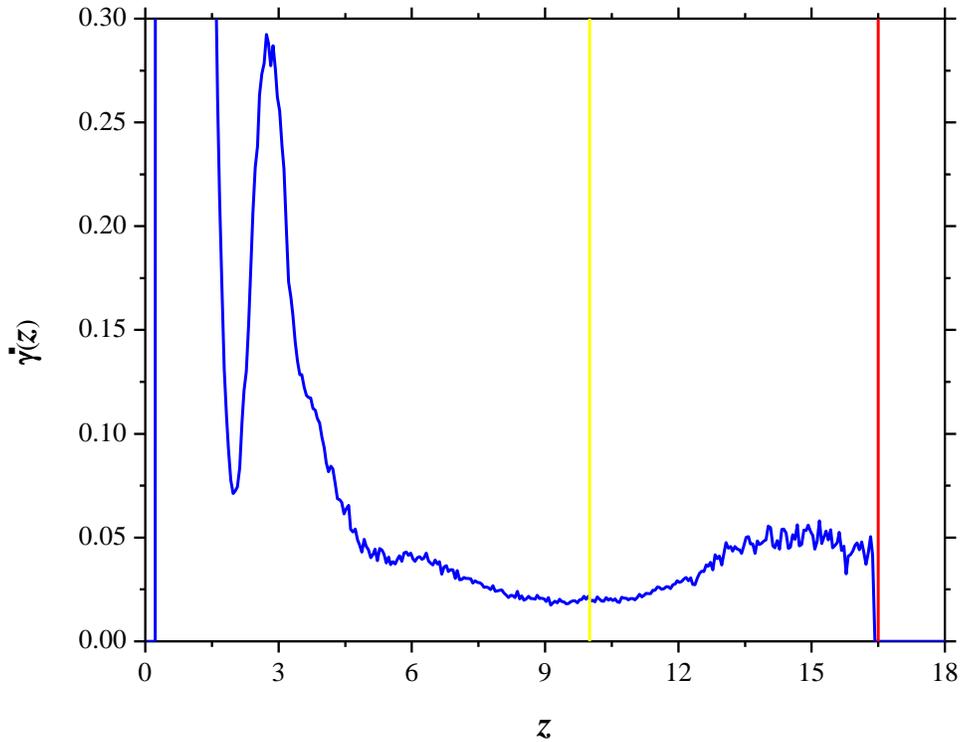

**Fig. 7** Profile of the average shear rate, $\dot{\gamma}$, experienced by the beads of all the brushes on the surface of the cancer cell. The vertical yellow and red lines indicate the position where the tip of the AFM probe is placed and where it ends, respectively. The vertical yellow and red lines indicate the position where the tip of the AFM probe is placed and where it ends, respectively. The axes are expressed in reduced DPD units.

Since the complex fluid made up of solvent particles and brush chains is under the influence of an external flow, there is a shear rate, which is defined as:

$$\dot{\gamma} = \frac{\partial v_x}{\partial z},\tag{5}$$

which is not constant, as in Couette flow between flat, parallel plates [26]. Although in the present case the separation between the AFM and the cell's surface is fixed, the brush is moving under the influence of an external harmonic force, hence the gradient in eq. (5) is not constant. However, in a thin slice along the $z$ – direction it is approximately constant and that allows one to construct a profile of shear rate as shown in Fig. 7. Two salient features of this profile are the oscillations close to the surface of the cell, as those in Fig. 6, which are the result of the magnitude of the collective average oscillations exerted on the surface of the cell as they propagate along the $z$ – direction; also, as the fluid approaches the surface of the AFM the shear rate profile becomes approximately constant because the oscillating nature of the external force is not dominant. Once the brushes reach the surface of the AFM they have more freedom to move and the shear rate increases accordingly.

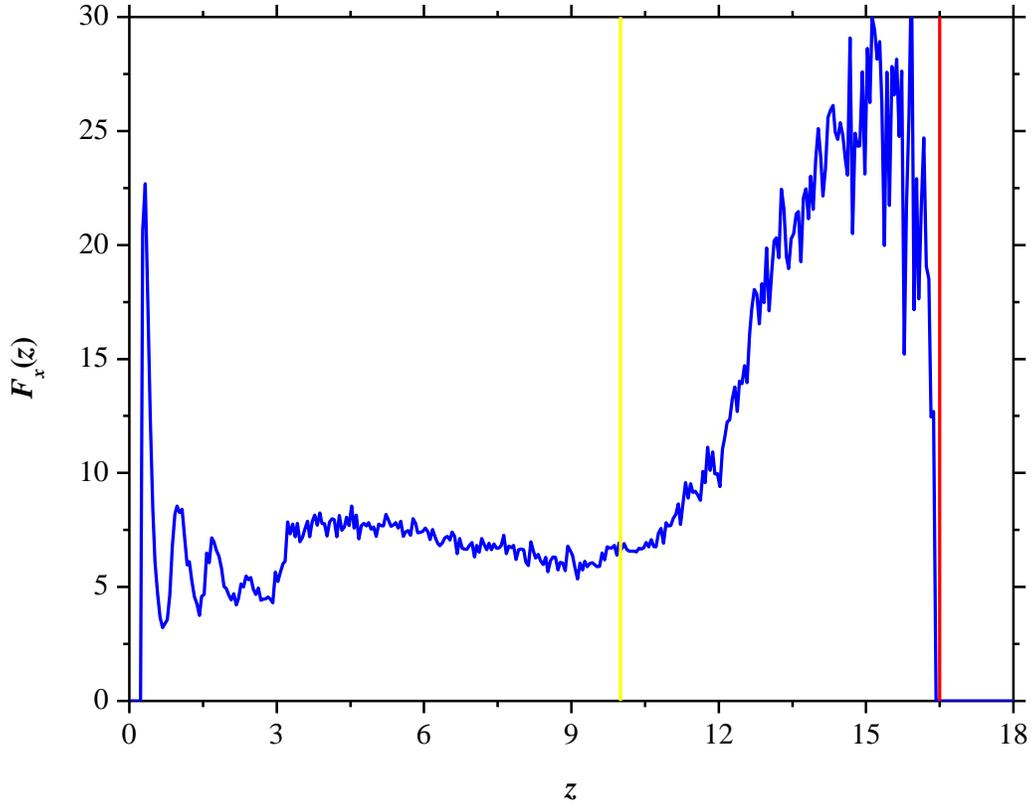

**Fig. 8** Average profile of the force applied on the $x$ – direction experienced by the brushes along the direction perpendicular to the plane of the cell on which the brushes are moving under the influence of an external oscillating force. The meaning of the vertical lines is the same as in the previous two figures. The scales on the axes are expressed in reduced DPD units.

Determining the shear rate is important because with it and with the response of the fluid to the average $x$ – component of the external force applied one can obtain the shear – dependent viscosity of the fluid, according to the relation

$$\eta = \frac{\langle F_x(\dot{\gamma})\rangle}{\dot{\gamma}} \quad . \tag{6}$$

The numerator in eq. (6) is shown in Fig. 8 as one moves along the direction perpendicular to the surface of the cell, while the denominator is shown in Fig. 7. The oscillations present in Fig. 8 close to the surface of the cell are particularly informative; they have approximately the same period while their amplitude is reduced as $z$ increases, effectively disappearing at distances larger than about $z = 3$ from the cell's surface. The oscillations are the collective response of the brushes to the externally applied oscillatory force on the plane of the surface, and their disappearance after $z = 3$ is consequence of the thickness of the brush made up of the shortest chains, see the blue line in Fig. 5(b). For the medium – sized and large chains, the oscillatory motion of the fluid made up of brushes and solvent particles is averaged out, and the $x$ – component of the force remains approximately constant until the brushes reach the tip of the AFM (yellow vertical line in Fig. 8). Once the brushes are beyond this point, the collisions between the brushes and the surface of the AFM probe increase the average value of the force experienced by DPD beads along the $x$ – direction.

Using eq. (6) and the results presented in Fig. 7 and 8 we calculated the profile of the shear dependent viscosity, which is shown in Fig. 9. Clearly, the fluid displays non – Newtonian behavior, as expected for a fluid as complex as the present one. In particular, the viscosity increases as the distance from the surface of the cell grows, with a local maximum appearing at $z \approx 2$ because the shear rate has a local minimum at the same point, see Fig. 7 and eq. (6).

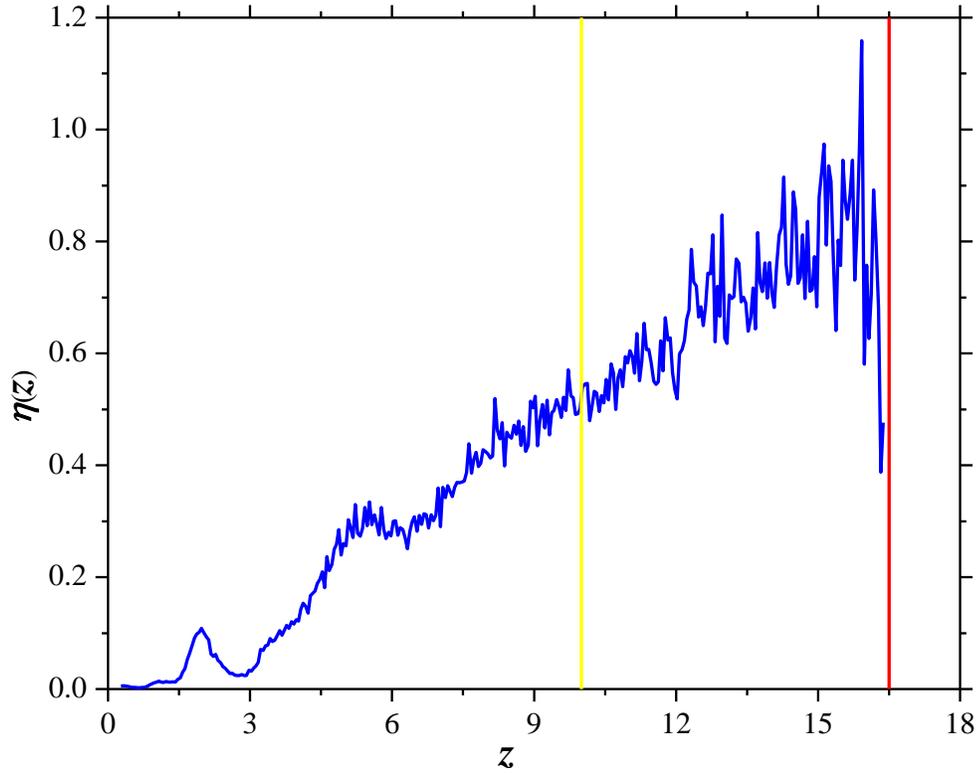

**Fig. 9** Profile of the shear dependent viscosity of the fluid made up of brushes and solvent particles, see eq. (6) and Figs. 7 and 8. The yellow and red vertical lines have the same meaning as in those figures. All quantities are expressed in reduced DPD units.

Beyond that point, the viscosity increases monotonically in a more or less linear fashion. This regime is the response of the fluid to the dominance of the force shown in Fig. 8, i.e., the collisions of the largest chains with the curved surface of the AFM probe. This is the first report of a viscosity profile in an AFM system that we are aware of. At distances close to the cell's surface, the fluid experiences an almost undamped response to the applied external force, which translates into an almost inviscid fluid response. However, as the particles move away from the surface, the influence of the oscillatory external force is damped and the increase in viscosity results from increased collisions between beads and the AFM probe.

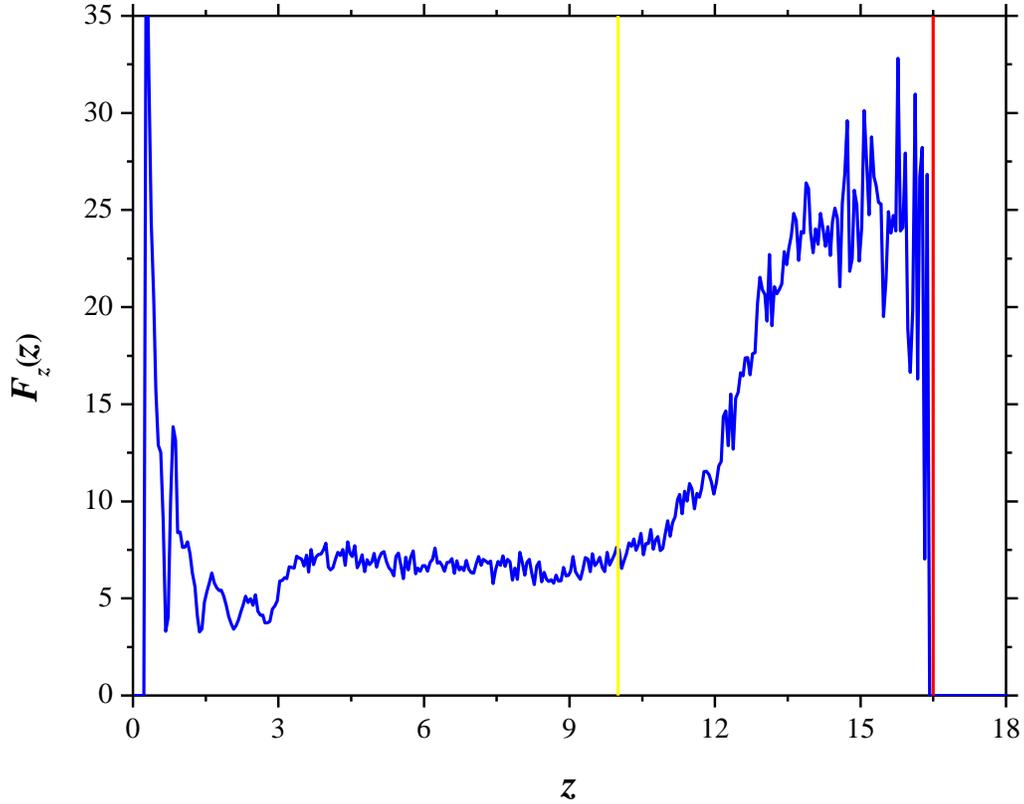

**Fig. 10** Profile of the average $z$ – component of the force experienced by the particles of the fluid as the distance from the surface of the cell ($z = 0$) is increased. Both axes are expressed in reduced DPD units. The vertical colored lines have the same meaning as in previous figures.

The $z$ – component of the force whose profile is shown in Fig. 10 is qualitatively very similar to its $x$ – component counterpart, shown in Fig. 8. The oscillation with decaying amplitude close to the cell's surface have the same origin as those in Fig. 8, i.e., they are the response of the shortest chains, mostly, which is why they disappear for $z > 3$. The almost constant profile in the region $3 \leq z \leq 10$ in Fig. 10 is indicative of the fact that there is no increment in the osmotic pressure in that region due to the oscillating motion of the surface of the cell. The value of the average $z$ – component of the force is about the same as the average $x$ – component of the force, in the same region (see Fig. 8), they have to do with the intrinsic properties of the fluid there, such as the concentration of particles and their interactions, but are insensitive to the influence of the external force.

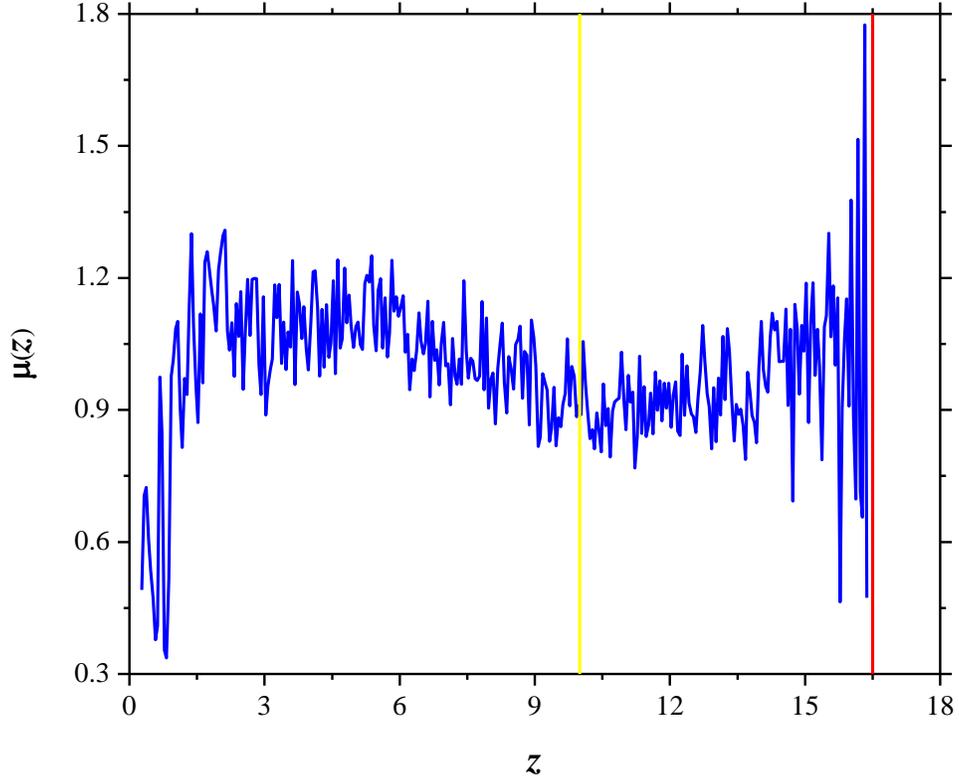

**Fig. 11** Friction coefficient profile, calculated according to eq. (7). It is a dimensionless number; the $x$ – axis is expressed in reduced DPD units. The vertical colored lines have the same meaning as in previous figures.

Finally, we calculate the friction coefficient along the $z$ – direction. This coefficient is defined as the ratio of the average value of the $x$ – component of the force on the chains, over the average value of its $z$ – component counterpart, see eq. (7):

$$\mu(z) = \frac{\langle F_x(z) \rangle}{\langle F_z(z) \rangle}, \qquad (7)$$

where the numerator is the quantity related to the $y$ – axis in Fig. 8 and the denominator corresponds to the $y$ – axis of Fig. 10. It is, by definition, a dimensionless number, which can be measured in experiments carried out with AFM [22], therefore it constitutes a very useful characteristic that allows direct comparison of our model with those experiments. The friction coefficient profile shown in Fig. 11 displays low values close to the surface of the cell, because the fluid is almost inviscid in that region, as one can also ascertain from Fig. 9. Then, as the distance from the cell increases, one sees a jump in µ to a value close to 1, because both force components have approximately the same value in a region where the shortest brush ends and the tip of the AFM appears. The slight oscillatory behavior of the friction coefficient seen in that interval is simply due to the same slight oscillations of the *x* – component of the force, see Fig. 8. Once the fluid reaches the probe of the AFM (the region between the yellow and red lines in Fig. 11), the average value of the friction coefficient increases, reaching a relatively large value (µ = 1.8) at the top of the simulation box. The values of the friction coefficient, seen in Fig. 11, are larger than those obtained with polymer brushes, using also DPD simulations [27, 28], because the proportion of solvent particles to brush beads is smaller in this work than in other reports. It is known that the solvent acts as a lubricant in polymer brushes [22, 23], and here we modeled a fluid formed predominately by brushes.

## IV Conclusions

The influence of an external oscillatory force on the surface of epithelial cells covered by brushes of different lengths, such as those observed in experiments carried out on human cervical cells with AFM, appears to be stronger on the smallest brush, which responds to the oscillations with small damping. However, for the largest brushes at the amplitude and frequency used in this work, the response appears to be slower, allowing for the relaxation of the chains and their averaged interactions. The profiles of the viscosity and friction coefficient, which are the first of their kind reported in the literature as far as we know, show that the interaction between the chains and the tip of the AFM is increased, as well as these rheological properties, not only because of their molecular characteristics such as density and length. But also, crucially, because the use of a three – body potential between bonds along the chains leads to an effectively larger persistence length of the brushes, which forces them to collide with the tip of the AFM and with each other. We believe this work is useful as a guide in the interpretation of recent experiments in fluids as complex as those dealt with here.

## Acknowledgements

AGG would like to thank S. J. Alas, S. Hernández, J. L. Menchaca and I. Sokolov for helpful discussions. JDHV and AGG acknowledge also the hospitality of E. Pérez and the Polymer Group at the Instituto de Física (UASLP), where this work was conceived.

## References

[1] E. L. Franco *et al*, *J. Infect. Dis*. **180**, 1415-1423 (1999). T Iftner *et al.*, *Br. J. Cancer* **88**, 1570 (2003).

[2] J. Guck *et al*. *Biophys. J*. **88**, 3689 (2005).

[3] M. Lekka *et al*., *Arch. Biochem. Biophys*. **518**, 151 (2012).

[4] S. E. Cross, Y. S. Jin, J. Rao, and J. K. Gimzewski, *Nature Nanotech*. **2**, 780 (2007).

[5] M. Plodinec *et al*., *Nature Nanotech*. **7**, 757 (2012).

[6] D. J. Müller & Y. F. Dufrêne, *Trends Cell Biol*. **21**, 461 (2011).

[7] Q. S. Li, G. Y. H. Lee, C. N. Ong, C. T. Lim, *Biochem. Biophys. Res. Comm*. **374**, 609 (2008).

[8] M. Lekka, *Nature Nanotech*. **7**, 691 (2012).

[9] T. M. Koch, S. Münster, N. Bonakdar, J. P. Butler, B. Fabry, *PLoS One* **7**, e33476 (2012).

[10] E. Jonietz, *Nature* **491**, S56 (2012).

[11] S. Iyer, R. M. Gaikwad, V. Subba-Rao, C. D. Woodworth, I. Sokolov, *Nature Nanotech*. **4**, 389 (2009).

[12] A. Gama Goicochea, S. J. Alas Guardado. *Sci. Rep.* **5**, 13218 (2015).

[13] Da-Kang Yao and Jin-Yu Shao, *Cell Mol Bioeng.* **1**, 75 (2008).

[14] P. J. Hoogerbrugge, J. M. V. A. Koelman, *Europhys. Lett.* **19**, 155 (1992).

[15] M. P. Allen, D. J. Tildesley, *Computer Simulation of Liquids*, Oxford University Press, New York (1989).

[16] P. Español, P. Warren, *Europhys. Lett.* **30**, 191 (1995).

[17] A. Gama Goicochea, M. A. Balderas Altamirano, R. López Esparza, M. A. Waldo, E. Pérez. *Eur. J. Phys.* **36**, 055032 (2015).

[18] C. Pastorino and A. Gama Goicochea, in *Selected Topics of Computational and Experimental Fluid Mechanics*, J. Klapp, G. Ruíz, A. Medina, A. López & L. Di G. Sigalotti (Eds.), Springer Book Series: Environmental Science and Engineering: Environmental Science, *Springer International Publishing Switzerland*, (2015).

[19] A. Gama Goicochea, M. A. Balderas Altamirano, J. D. Hernández and E. Pérez. *Comp. Phys. Comm.* **188**, 76 (2015).

[20] T. Murtola, A. Bunker, I. Vattulainen, M. Deserno, and M. Karttunen, *PCCP* **11**, 1869 (2009).

[21] I. Vattulainen, M. Karttunen, G. Besold, and J. M. Polson, *J. Chem. Phys.* **116**, 3967 (2002).

[22] J. Klein *et al.*, *Nature* **370**, 634 (1994).

[23] E. Eiser and J. Klein, *Macromolecules*, **40**, 8455 (2007).

[24] A. Gama Goicochea, M. Romero-Bastida, R. López-Rendón, *Mol. Phys.* **105**, 2375 (2007).

[25] M. E. Velázquez and A. Gama Goicochea, M. González-Melchor, M. Neria, and J. Alejandre, *J. Chem. Phys.* **124**, 084104 (2006).

[26] C. W. Macosko, *Rheology Principles, Mesurements, and Applications*, Wiley-VCH Inc, New York (1994).

[27] F. Goujon, P. Malfreyt and D. J. Tildesley, *Soft Matter* **6**, 3472 (2010).

[28] A. Gama Goicochea, E. Mayoral, J. Klapp, and C. Pastorino. *Soft Matter* **10**, 166 (2014).